\documentclass[aps,prb,twocolumn,superscriptaddress,showpacs,amsmath,amssymb,longbibliography]{revtex4-2}
\usepackage{graphicx}
\usepackage[colorlinks=true,citecolor=blue,linkcolor=blue,urlcolor=blue,bookmarks=true]{hyperref}

\usepackage[T1]{fontenc}
\usepackage[utf8]{inputenc}
\usepackage[english]{babel}
\usepackage{amsmath}
\usepackage{xcolor}
\usepackage{soul}
\usepackage{physics}

\definecolor{nred} {RGB}{224,0,0}
\definecolor{nblue} {RGB}{28,130,185}
\definecolor{dgreen}{RGB}{78,138,21}
\definecolor{norange}{RGB}{230,120,20}

\begin{document}

\title{Peak of fidelity susceptibility is controlled by stiffness of perturbation}
\author{J. Paw\l{o}wski}
\affiliation{Institute of Theoretical Physics, Faculty of Fundamental Problems of Technology, Wroc\l{a}w University of Science and Technology, 50-370 Wroc\l{a}w, Poland}  
\author{L. Vidmar}
\affiliation{Department of Theoretical Physics, J. Stefan Institute, SI-1000 Ljubljana, Slovenia}
\affiliation{Department of Physics, Faculty of Mathematics and Physics, University of Ljubljana, SI-1000 Ljubljana, Slovenia\looseness=-1}
\author{M. Mierzejewski}
\affiliation{Institute of Theoretical Physics, Faculty of Fundamental Problems of Technology, Wroc\l{a}w University of Science and Technology, 50-370 Wroc\l{a}w, Poland}
\author{P. Łydżba}
\affiliation{Institute of Theoretical Physics, Faculty of Fundamental Problems of Technology, Wroc\l{a}w University of Science and Technology, 50-370 Wroc\l{a}w, Poland}

\begin{abstract}
The fidelity susceptibility quantifies the sensitivity of energy eigenstates to perturbations. It can develop a peak signaling the breaking of integrability and the onset of ergodic behavior. It was proposed that this peak exhibits maximal divergence with system size for integrability-preserving perturbations, a phenomenon known as maximal chaos, while it is absent for integrability-breaking ones. However, these two classes of perturbations have never been sharply defined, and examples that challenge this classification have also been reported in the literature. Here, we provide a simple argument showing that the height of the peak is proportional to the stiffness of a perturbation in the integrable limit. Since the stiffness corresponds to the mean of squared diagonal matrix elements, its value is determined by the projection of a perturbation onto local integrals of motion. We establish that maximal chaos is observed for local perturbations, for which this projection does not vanish in the thermodynamic limit. We further demonstrate that other local perturbations can exhibit arbitrary intermediate behaviour between maximal divergence and absence of a peak. We support our argument with numerical simulations of the XXZ spin chain with next-nearest-neighbor interactions, using a systematically constructed set of local perturbations.
\end{abstract}

\maketitle

\section{Introduction}

Closed quantum systems are generically ergodic, meaning that they thermalize after being driven out of equilibrium~\cite{Srednicki_1999,deutsch18,rigol08,polkovnikov2011,D_Alessio_2016}. Consequently, the details of the initial state that can be extracted locally are lost at long times, apart from the expectation values of a few local integrals of motion (LIOMs), like energy and particle number. This delocalization of information poses challenges for many applications, such as quantum memory, and has motivated considerable interest in integrable systems. These systems possess an extensive number of LIOMs, which constrain their dynamics and limit information scrambling~\cite{PhysRevLett.98.050405,bertini15,Vidmar_2016,He_2026}. Other nonergodic systems, such as many-body localized~\cite{Nandkishore_2015,RevModPhys.91.021001,Sierant_2025,PhysRevE.102.062144} or fragmented~\cite{PhysRevB.101.174204,PhysRevX.10.011047,Rakovszky_2020,Moudgalya_2021,Moudgalya_2022} systems, have also attracted considerable attention. However, their properties are still debated, so we do not focus on them in this study.

It is believed that in the thermodynamic limit, an arbitrarily small perturbation breaks integrability and induces ergodic behavior~\cite{PhysRevB.69.054403,Modak_2014,PhysRevB.90.075152,PhysRevE.81.036206,PhysRevB.105.214308}. Still, finite systems can exhibit signatures of integrability in the presence of perturbations of nonzero strength. Numerous quantities have been used to investigate the crossover from integrable to ergodic behavior, including spectral statistics~\cite{Brody1973,Prosen1994,Szasz_2021} and entanglement entropy~\cite{Beugeling_2015,LeBlond_2019,PhysRevE.109.024117}. These quantities, however, typically change monotonically, and often in a nonuniversal manner, between the two regimes, making it difficult to identify a unique crossover point. In contrast, the fidelity susceptibility, or the closely related adiabatic gauge potential, develops a peak signaling the change from integrable to ergodic behavior~\cite{LeBlond_2021,Ren_2022,Najafzadeh_2026,Kim_2026b}.

The fidelity susceptibility quantifies the sensitivity of energy eigenstates to perturbations~\cite{Kolodrubetz_2017,PhysRevB.109.195162,Kim_2026}. Since it is defined through the ratio of squared matrix elements of perturbation to squared energy differences, it is highly sensitive to the hybridization of nearby energy eigenstates. This often enables a faster detection of changes in the energy eigenstate properties compared with other measures~\cite{PhysRevX.10.041017,PhysRevB.111.184203,Lisiecki_2025, Tokarczyk_2026}. Additionally, it is related to the Fourier transform of the (connected) autocorrelation function~\cite{PhysRevX.10.041017}. This allows us to introduce a frequency cutoff into the definition of fidelity susceptibility and look for signatures of integrability at a selected timescale~\cite{Karve_2025,Author2026}. This is particularly relevant for experiments, for which observing integrable behavior up to a certain time, often much shorter than the longest physical timescale, is often sufficient~\cite{PhysRevX.7.041062,Jin_2025}. Finally, the fidelity susceptibility may be relevant in the context of initial-state preparation~\cite{Sugiura_2021} and has recently been invoked in the studies of the robustness of Hamiltonian learning~\cite{Gjonbalaj}.

Nevertheless, there remains a puzzle concerning the peak of fidelity susceptibility. It is not particularly surprising that its height, and thus the sensitivity of energy eigenstates, depends on the choice of perturbation. It was proposed that the divergence of the peak with system size is maximal when the perturbation is integrability-preserving, while the peak does not appear when the perturbation is integrability-breaking~\cite{PhysRevX.10.041017}. However, these two classes of perturbations have never been sharply defined. Integrability-preserving perturbations may be understood as those that either commute with the Hamiltonian in the integrable limit or allow one to move along a continuous family of integrable Hamiltonians. Nevertheless, examples that challenge this classification have been reported in the literature. For example, it has been demonstrated that the behavior of fidelity susceptibility barely differs between integrability-preserving and integrability-breaking perturbations in the XXZ spin chain with next-nearest-neighbor interactions~\cite{LeBlond_2021}. We note that an analogous observation has been reported in the Hilbert-space fragmented systems~\cite{Lisiecki_2025}, though such systems are beyond the scope of the present study.

In this manuscript, we provide a simple analytical argument showing that the height of the peak is determined by the stiffness of the perturbation at the integrable point. Since the stiffness corresponds to the mean of squared diagonal matrix elements, its value is determined by the projection of the perturbation onto LIOMs~\cite{Suzuki_1971,PhysRevLett.114.140601,PhysRevLett.124.040603}. Consequently, maximal chaos is observed for local perturbations with a nonvanishing projection. Most importantly, we demonstrate that one may choose a local perturbation that produces an arbitrary intermediate behavior ranging from maximal divergence to the complete absence of a peak. We support our analytical argument with numerical simulations of the XXZ spin chain with next-nearest-neighbor interactions. Instead of studying arbitrarily selected local perturbations, we systematically generate  an orthonormal set of operators using the numerical procedure from Ref.~\cite{Paw_owski_2025}, which works directly in the thermodynamic limit.

The manuscript is organized as follows. In Sec.~\ref{sec:fidelity}, we define the fidelity susceptibility and present our analytical argument. In Sec.~\ref{sec:model_method}, we introduce the considered model, the XXZ spin chain with next-nearest-neighbor interactions, and describe the procedure for generating an orthonormal set of local perturbations. In Sec.~\ref{sec:numerical} we discuss the numerical results, first for the generated orthonormal set of perturbations in Sec.~\ref{sec:basis}, and then for generic local operators in Sec.~\ref{sec:generic}. Finally, we investigate local perturbations with zero stiffness and relate their existence to symmetry constraints in Sec.~\ref{sec:zeros}. We summarize our findings in Sec.~\ref{sec:conclusions}.

\section{Fidelity susceptibility}
\label{sec:fidelity}

In this manuscript, we study how the choice of local perturbation influences the behavior of the fidelity susceptibility across the integrability-breaking transition. To this end, we consider the Hamiltonian, $H(\Delta_2)$, which is integrable at $\Delta_2=0$. Increasing the integrability-breaking strength, $\Delta_2$, drives the system towards ergodicity [see Eq.~\eqref{eq:xxz}]. Since the fidelity susceptibility characterizes the sensitivity of energy eigenstates to perturbations of the Hamiltonian, we introduce a perturbed model,
\begin{equation}
H(\Delta_2,\gamma)=H(\Delta_2)+\gamma O,
\end{equation}
where $O$ is a local perturbation, which means that it can be written as $O=\sum_l o_l$, where the sum runs over all spatial positions, and each $o_l$ is supported on a subsystem of size $L_s$, which is smaller than the total system size $L$. To distinguish between the two, we also refer to $O=o_l$ as an ultra-local perturbation. To simplify the presentation, we also assume that $O$ is Hermitian, traceless, and has unit Hilbert-Schmidt norm, i.e., $\text{Tr}\, O=0$ and $||O||^2={\cal D}^{-1}\text{Tr}[O^2]=1$, where $\cal{D}$ is the Hilbert space dimension. The eigenstates and eigenvalues of $H(\Delta_2,\gamma)$ are defined by 
\begin{equation}
    H(\Delta_2,\gamma)\ket{n(\Delta_2,\gamma)}=E_n(\Delta_2,\gamma)\ket{n(\Delta_2,\gamma)}.
\end{equation} 
The overlap between the eigenstates at $\gamma$ and $\gamma+d\gamma$, to the leading order in $d\gamma$, is given by~\cite{PhysRevLett.99.100603,PhysRevE.76.022101},
\begin{equation}
|\langle n(\Delta_2,\gamma+d\gamma)|n(\Delta_2,\gamma)\rangle|^2
\approx 1-\chi_n d\gamma^2,
\end{equation}
where
\begin{equation}
\label{eq:fs_n}
    \chi_n(\Delta_2,\gamma)=\sum_{m\neq n}\frac{|\langle n(\Delta_2,\gamma)|O|m(\Delta_2,\gamma)\rangle|^2}{\omega_{nm}^2(\Delta_2,\gamma)}
\end{equation}
stands for the fidelity susceptibility, while $\omega_{nm}(\Delta_2,\gamma)=E_n(\Delta_2,\gamma)-E_m(\Delta_2,\gamma)$ is the energy difference. Since we are interested in how the sensitivity of the eigenstates of the original Hamiltonian, $H(\Delta_2)$, changes with the integrability-breaking parameter, $\Delta_2$, we evaluate the fidelity susceptibility, $\chi_n(\Delta_2,\gamma)$, at  vanishing perturbation, $\gamma\to0$. In what follows, we suppress the explicit dependence on $\Delta_2$ and $\gamma$. We also denote the matrix elements as $O_{nm}=\langle n|O|m\rangle$.

We now introduce some quantities that are convenient in the integrable limit, i.e., $\Delta_2=0$. First, we consider a set of LIOMs, $I_j$, assumed to be Hermitian, traceless, and orthonormal with respect to the Hilbert--Schmidt inner product, i.e., $\langle I_i I_j\rangle=\mathcal{D}^{-1}\mathrm{Tr}[I_i I_j]=\delta_{ij}$. We also introduce the stiffness, defined as the mean of the squared diagonal matrix elements,
\begin{equation}
\label{eq:stiff}
    \sigma=\frac{1}{\mathcal D}\sum_n O_{nn}^2,
\end{equation}
where, for simplicity, we assumed that the energy spectrum is nondegenerate. The stiffness is determined by the projections of $O$ onto the LIOMs, i.e., $d_j=\langle OI_j\rangle$. To see this, we decompose $O$ as
\begin{equation}
    O=\sum_j d_j I_j+O_\perp,
\end{equation}
where $O_\perp$ is orthogonal to all LIOMs. Substituting this decomposition into the definition of stiffness yields
\begin{equation}
\label{eq:stiff2}
    \sigma=\sum_j d_j^2+
    \frac{1}{\mathcal D}\sum_n[O_\perp]_{nn}^2
    \approx\sum_j D_j,
\end{equation}
where we have introduced $D_j=d_j^2$. We have also used the fact that $\mathcal{D}^{-1}\sum_n [I_j]_{nn} A_{nn}=0$ for any operator $A$ orthogonal to a LIOM $I_j$, and that the stiffness of $O_\perp$ is small and quickly decreases with system size~\cite{PhysRevLett.124.040603}. We return to the latter point later in this section.

To proceed, we note that both integrable~\cite{Brenes_2020b,Brenes_2020,PhysRevE.102.062113} and ergodic~\cite{Srednicki_1999,Sch_nle_2021} systems satisfy the off-diagonal part of the Eigenstate Thermalization Hypothesis (ETH) ansatz, i.e.,
\begin{equation}
\label{ETH}
O_{nm}=\rho(E_{nm})^{-\frac{1}{2}}f(E_{nm},\omega_{nm})R_{nm},
\end{equation}
where $E_{nm}=(E_n+E_m)/2$ is the mean energy, $\rho(E_{nm})$ is the density of states, $R_{nm}$ is a random variable with zero mean and unit variance, and $f(E_{nm},\omega_{nm})$ is the envelope function, which is smooth in both of its arguments. The latter function is related to the Fourier transform of the (connected) autocorrelation function. This can be readily seen from a simple calculation:
\begin{equation}
\begin{aligned}
& \int\limits_{-\infty}^{\infty} dt\;
\frac{e^{-i\omega t}}{2\pi}C_\infty(t)=\frac{1}{\mathcal D}\sum_{n,m\neq n}
\int\limits_{-\infty}^{\infty} dt\;
\frac{e^{-i(\omega-\omega_{nm})t}}{2\pi}
|O_{nm}|^2\\
&= \frac{1}{\mathcal D}
\sum_{n,m\neq n}|O_{nm}|^2
\delta(\omega-\omega_{nm})
\approx |f(\overline E,\omega)|^2,
\end{aligned}
\end{equation}
where $C_\infty(t)=\frac{1}{\mathcal D}
\sum_n[\langle n|O(t)O|n\rangle
-\langle n|O|n\rangle^2]$, $\delta(\omega)$ is the Dirac delta and $\overline E={\cal D}^{-1}\sum_{n} E_n$.
We emphasize that the key differences between integrable and ergodic systems are encoded in the statistical properties of $R_{nm}$ and in the behavior of $f(E_{nm},\omega_{nm})$. Specifically, $R_{nm}$ are Gaussian distributed in ergodic systems, while in integrable systems, their distribution develops long tails and can be to a certain degree approximated by a log-normal distribution~\cite{LeBlond_2019,PhysRevB.110.134206}. We discuss the behavior of $f(E_{nm},\omega_{nm})$ below.

The sum in Eq.~\eqref{eq:fs_n} is dominated by the contribution with the smallest energy difference. For a typical fidelity susceptibility, $\chi_{\rm typ} = \exp(\overline{\ln \chi_n})$, this corresponds to $\omega_{nm}\approx\omega_H$ near $E_{nm}\approx \overline{E}$, where the overline denotes the averaging over energy eigenstates, while $\omega_H$ denotes the typical level spacing.
For a system with short-range interactions, one finds that
$\omega_H\approx a \sqrt{L}/\cal{D}$, where $a=const$. This allows us to write
\begin{equation}
\label{eq:fs_typ}
\chi_\text{typ}
\approx \frac{|f(\overline{E},\omega_H)|^2}{\omega_H},
\end{equation}
where we have used that $|R_{nm}|^2\approx 1$ and $\rho(\overline{E})\approx\omega_H^{-1}$.

In ergodic systems, $|f(\overline{E},\omega)|^2$ develops a plateau as $\omega\rightarrow\omega_H$~\cite{D_Alessio_2016}. For this reason, $\chi_\text{typ}\sim\omega_H^{-1}$, and so it grows exponentially with system size. In contrast, in integrable systems, $|f(\overline{E},\omega)|^2$ is suppressed at small energy differences, or even develops a gap for finite systems~\cite{Brenes_2020b,Brenes_2020,PhysRevE.102.062113}. This leads to a slower finite-size scaling of $\chi_\text{typ}$~\cite{PhysRevX.10.041017,Lisiecki_2025}.

The evolution of $|f(\overline{E},\omega)|^2$ across the integrability-breaking transition has been previously described in Ref.~\cite{PhysRevLett.127.230603}. In the integrable limit, the envelope function develops a Drude-like peak,
\begin{equation}
\label{eq:integrable}
|f(\overline{E},\omega\approx\omega_H)|^2\approx\sum_j D_j\delta(\omega).
\end{equation}
Simultaneously, once integrability is broken, $I_j$ are no longer conserved and acquire finite relaxation times, $\tau_j$. This leads to the broadening of the Drude-like peak. In other words, spectral weight is redistributed from \(\omega=0\) to small but nonzero energy differences, leading to the formation of a peak in the fidelity susceptibility. Arguments along these lines have already been discussed in Ref.~\cite{Tokarczyk_2026}. In this case, the envelope function can be approximated as
\begin{equation}
\label{eq:nearly}
|f(\overline{E},\omega\approx\omega_H)|^2\approx\sum_j \frac{D_j}{\pi}\frac{\tau_j}{(\omega\tau_j)^2+1},
\end{equation}
where we have assumed Lorentzian broadening. Although the expressions from Eqs.~\eqref{eq:first}--\eqref{eq:last} depend quantitatively on the choice of broadening, we expect their qualitative behavior to be independent of it. Additionally, typically a single integral of motion, $I_k$, makes the dominating contribution to Eqs.~\eqref{eq:integrable} and~\eqref{eq:nearly}
so that the sum, $\sigma=\sum_{j}D_j$, is  of the same order of magnitude as the largest summand, $D_k=\text{max}[D_j]$.

Taking into account that the emergence of the peak is expected for $\tau_k\approx\omega_H^{-1}$, and marking the corresponding integrability-breaking strength as $\Delta_2=\Delta_2^*$, we can write 
\begin{equation}
\label{eq:first}
    |f(\overline{E},\omega=\omega_H)|^2\approx \frac{\sigma}{2\pi\omega_H}.
\end{equation}
Consequently, the height of the peak of the fidelity susceptibility $\chi_\text{typ}^*$, emerging at $\Delta_2=\Delta_2^*$, appears to be determined by the stiffness of the perturbation in {\it the integrable limit},
\begin{equation}
\label{eq:final}
    \chi_\text{typ}^*\approx\frac{\sigma}{2\pi\omega_H^2}\approx \frac{{\cal D}^2}{2\pi a^2 L}\sigma\,,
\end{equation}
where we used that $\omega_H = a\sqrt{L}/{\cal D}$.
Consequently, $\chi_\text{typ}^*$ is determined by how strongly $O$ projects onto the LIOMs.

In general, one can fine-tune an operator for which all local integrals of motion equally contribute to the stiffness. In such a case, one can derive a useful bound relating the stiffness to the fidelity susceptibility. We recall that 
an integrable system of size $L$ hosts $bL$ local integrals of motion, where $b$ is a constant of the order of unity. Since all of them can contribute to Eqs.~\eqref{eq:integrable} and~\eqref{eq:nearly}, we find the following bound:
\begin{eqnarray}
\label{eq:last}
&&\sum_{j=1}^{bL} \frac{D_j}{\pi}\frac{\tau_j}{(\omega_H\tau_j)^2+1}
\le
bL {\rm max} \left( \frac{D_j}{\pi \omega_H}\frac{\omega_H \tau_j}{(\omega_H\tau_j)^2+1} \right) \nonumber \\
&& \le bL \frac{D_k}{2 \pi \omega_H } \le bL \frac{\sigma}{2\pi \omega_H}.
\label{eq:bound}
\end{eqnarray}
Consequently, operators with small \(\sigma\) cannot cause large peaks in the fidelity susceptibility, as long as the picture of Drude-like peak  broadening remains valid.

The maximal divergence of the typical fidelity susceptibility, $\chi_\text{typ}\sim \omega_H^{-2}$, is known as maximal chaos and has been proposed to occur for integrability-preserving perturbations. These include integrals of motion, such as $O=I_k$~\cite{PhysRevX.10.041017}, as well as certain local perturbations that allow one to move along a continuous family of integrable Hamiltonians, such as the nearest-neighbor kinetic energy and interactions in the case of the XXZ chain. However, Eq.~\eqref{eq:final} indicates that the class of perturbations leading to this maximally divergent behavior is broader. It is sufficient that the projection of $O$ onto $I_k$, and so the stiffness $\sigma$, is independent of the system size. This occurs for an arbitrary operator of the form $O=\sum_{j}d_jI_j+O_\perp$ with at least one $d_j\neq 0$.

Additionally, it has previously been conjectured that the typical fidelity susceptibility fails to develop a peak near the transition for integrability-breaking perturbations~\cite{PhysRevX.10.041017}, although this conclusion has been questioned in other studies~\cite{LeBlond_2021,Lisiecki_2025}. Note that the expression from Eq.~\eqref{eq:final} can be extended to include all conserved quantities, not only local ones. Their contribution is usually disregarded, since the projections of a local perturbation onto them vanish in the thermodynamic limit~\cite{PhysRevLett.124.040603}. However, if a local perturbation has a dominant projection onto a product of LIOMs, the stiffness can decay polynomially with system size, i.e., $\sigma\sim L^{-\kappa}$~\cite{PhysRevLett.124.040603}. As a result, the typical fidelity susceptibility can exhibit an intermediate behavior between maximal divergence and absence of a peak, i.e., $\chi_\text{typ}\sim \omega_H^{-2}L^{-\kappa}$. The interpretation of this result depends on how one defines local observables. If they are required to be supported on a subsystem whose size, $L_s$, is independent of the total system size, $L$, then projections onto products of LIOMs give rise to polynomial corrections to maximal chaos. However, even in such a case, these corrections can be non-negligible for systems accessible to numerical simulations. This is demonstrated in Sec.~\ref{sec:numerical}, where we classified observables as local if their support $L_s$ is finite and does not exceed half the size of the smallest system studied. On the other hand, if the only requirement is that $L_s=aL^b\ll L$, allowing $b>0$, then deviations from maximal chaos can be significant for arbitrarily large systems.

Finally, for local perturbations that are orthogonal to both LIOMs and their products, the stiffness decays exponentially with system size, i.e., $\sigma\sim\omega_H$. This leads to the typical fidelity susceptibility having the same scaling near $\Delta_2=\Delta_2^*$ as deep in the ergodic regime, $\chi_\text{typ}\sim \omega_H^{-1}$, and hence to the absence of a peak. Although finding such a local perturbation requires fine-tuning, it is not impossible, as discussed in Sec.~\ref{sec:zeros}.

Note that the derivation of $\chi_\text{typ}$ in Eq.~\eqref{eq:final} is based on the assumption that breaking integrability leads to the broadening of the Drude-like peak. The peak is then formed when this broadening becomes of the order of the typical level spacing. Nevertheless, when a local perturbation has zero stiffness in the integrable limit, i.e., $O_{nn}=0$, no broadening can occur. In such a case, the envelope function is insensitive to the value of $\Delta_2$ and, so, the typical fidelity susceptibility fails to develop a peak. This is analyzed in detail in Sec.~\ref{sec:zeros}. Moreover, this relation fails when the considered local perturbation has a nonzero projection onto a LIOM that remains conserved for arbitrary $\Delta_2$. In particle systems, such a LIOM is typically the total particle number, while in spin systems it is the total magnetization. We also note that the expression for $\chi_\text{typ}$ in Eq.~\eqref{eq:final} can be obtained by relaxing the condition of a dominant projection while requiring all relevant LIOMs to have similar relaxation times, i.e., $\forall_j \tau_j\approx \omega_H^{-1}$. Finally, it is worth emphasizing that $\sigma\sim L^{-1}$ can arise not only when a local perturbation has a projection onto a product of LIOMs, but also when an ultra-local perturbation has a projection onto a LIOM.

Before moving on, it is worth mentioning that maximal chaos has recently been explained in terms of Fermi's golden rule (FGR), see Refs.~\cite{4l42-l7pk, PhysRevB.111.184203,Kim_2026b}. This approach considers the regularized version of fidelity susceptibility, whose behavior can be approximated as $\chi_\text{typ}(\mu)\approx |f(\overline{E},\mu)|^2/\mu$ [compare with Eq.~\eqref{eq:fs_typ}], where $\mu$ is an energy cutoff. The argument assumes that breaking integrability leads to a Lorentzian broadening of the envelope function, $|f(\overline{E},\omega)|^2$, with a width determined by the FGR relaxation rate, $\Gamma$. The peak position is then determined by $\Gamma=\mu$, yielding $\chi_\text{typ}[\mu]\sim\mu^{-2}$. For LIOMs, it seems natural to identify the FGR relaxation rate with the relaxation rate of the observable itself. However, for more general observables, such as those considered in this manuscript, no simple relation between these relaxation rates is guaranteed.

\section{Model and operators}
\label{sec:model_method}

We consider the XXZ chain with the next-nearest-neighbor interactions and the periodic boundary conditions,
\begin{equation}
\label{eq:xxz}
\begin{split}
H & =\frac{1}{2} \sum_{l=1}^{L} \left( S^+_l S^-_{l+1}+S^-_l S^+_{l+1}\right)+ \Delta  \sum_{l=1}^{L}  S^z_l S^z_{l+1}\\
& +\Delta_2  \sum_{l=1}^{L}  S^z_l S^z_{l+2},
\end{split}
\end{equation} 
which for $\Delta_2=0$ is a paradigmatic integrable model~\cite{yang66,yang66b,yang66c,Prosen_2011,Pereira_2014}. In the above equation, $S_l^\alpha$ are the spin-$1/2$ projection operators along axis $\alpha=x,y,z$ at site $l=1,...,L$, while $S_l^{\pm}=S_l^x\pm iS_l^y$ are the spin-flip operators at the same site. The Hamiltonian from Eq.~\eqref{eq:xxz} conserves the total spin projection, $S^z=\sum_l S_l^z$, and is translationally invariant. Additionally, it respects mirror symmetry and, thus, exhibits degeneracies. Two degenerate energy eigenstates either belong to different momentum sectors or both belong to $k=0$ or $k=\pi$ momentum sectors. Throughout the manuscript, we fix $\Delta=3/4$.

We do not want to study arbitrarily selected local operators. Instead, we generate an orthonormal set according to the numerical procedure from Ref.~\cite{Paw_owski_2025}. Simple codes implementing this numerical procedure are available in a public repository~\cite{Pawlowski_2025_github}. It begins by constructing
\begin{equation}
A=\sum_{l=1}^{L} a_l a_{l+1}\dots a_{l+M-1},
\end{equation}
where, for each site $i=l,\dots,l+M-1$, we select $a_i$ from the set of operators $\{S_i^+,S_i^-,S_i^z,I_i\}$, with $I_i$ corresponding to the identity operator.
Consequently, $M$ denotes {\it the maximal number} of consecutive sites on which $A$ can act nontrivially, defining its support. Hermitian operators can be formed as two orthogonal combinations, $A_\text{real}=A+A^\dagger$ and $A_\text{imag}=i(A-A^\dagger)$. We restrict considerations to $A_\text{imag}$, as they are structureless, i.e., they do not project onto the Hamiltonian. From all possible $A_\text{imag}$, we retain only those that commute with $S^z$. This means that the numbers of $S_i^+$ and $S_i^-$ are forced to be equal. Finally, we normalize $A_\text{imag}$ to unit Hilbert-Schmidt norm, which is achieved numerically. We refer to the resulting set of orthonormal local operators as $B^s$, where $s$ denotes the operator index.

Finally, we numerically solve the following eigenproblem in the full Hilbert space of dimension $2^L$:
\begin{equation}
\label{eq:eigen}
    \sum_{sp} V_{\alpha s} F_{sp} (V^T)_{p\beta}=\lambda_\alpha \delta_{\alpha\beta},
\end{equation}
where $F_{sp}=\langle f(B^s)f(B^p)\rangle$, with $f(B^s)=i[B^s,H]$, while $V_{sp}$ is an orthogonal matrix. We emphasize that $F_{sp}$ for finite $M$ can be calculated exactly in \textit{the thermodynamic limit} using only the operator algebra, as implemented in the software package from Ref.~\cite{Loizeau_2025}. Since $f$ is linear, the eigenproblem can be rewritten as
\begin{equation}
    \langle f(O^\alpha)f(O^\beta)\rangle=\lambda_\alpha\delta_{\alpha\beta}\,,
\end{equation}
with
\begin{equation} \label{def:Oalpha}
    O^\alpha=\sum_{s}V_{\alpha s} B^s.
\end{equation}
The operators $O^\alpha$ are local and form an orthonormal set, $\langle O^\alpha O^\beta\rangle=\delta_{\alpha\beta}$, which we will consider in the numerical simulations in the following sections. We note that $||f(O^\alpha)||^2=\lambda_\alpha$, so that each $\lambda_\alpha$ is non-negative and determines the squared Hilbert-Schmidt norm of $i[O^\alpha,H]$. For simplicity, we refer to $\lambda_\alpha$, rather than its square root, as the commutator norm from now on. Two conclusions follow from this observation: (i)~$\lambda_\alpha=0$ corresponds to a LIOM, while (ii)~the inverse square root of $\lambda_\alpha\neq0$ provides a crude approximation of the relaxation time~\cite{PhysRevB.96.214301,PhysRevB.97.094206,Kim_2015}.

It has been demonstrated that the described algorithm generates all LIOMs~\cite{Paw_owski_2025}. Additionally, due to the orthogonality of the set, $O^\alpha$ with $\lambda_\alpha\neq 0$ do not have projections onto LIOMs. However, in a finite system, they can have a non-zero projection onto products of LIOMs. In other words, they can have a polynomially decaying stiffness, see the discussion in Sec.~\ref{sec:fidelity}. We analyze the behavior of the typical fidelity susceptibility for this set in Sec.~\ref{sec:basis}. Since the prediction for the scaling of the peak from Eq.~\eqref{eq:final} assumes that a local perturbation has dominant projection onto a single conserved quantity (LIOM or a product of LIOMs), we expect it to be valid in the limit $\sigma\to 1$. Furthermore, in Sec.~\ref{sec:generic}, we consider linear combinations of two local perturbations from this set. We confirm that, in this case, the prediction from Eq.~\eqref{eq:final} remains valid even for $\sigma\ll 1$, provided that at least one of the constituent local perturbations has a large stiffness $\sigma\gtrsim 0.1$.

\section{Results}
\label{sec:numerical}

The numerical results presented in this section are obtained in the symmetry sector with a fixed total magnetization, $S^z=0$. Since the considered operators, $O^\alpha$, are translationally invariant, they have no matrix elements between energy eigenstates belonging to different momentum sectors. This allows us to perform the numerical calculations of $\chi_\text{typ}$ and $\sigma$ separately in each momentum sector and then average the results over all sectors. To avoid dealing with degeneracies, we do not consider the $k=0$ and $k=\pi$ momentum sectors. When calculating $\chi_\text{typ}$, we consider $40\%$ of the eigenstates in the middle of the spectrum. For this reason, the quantity that we refer to as the Hilbert-space dimension, $\mathcal{D}$, is actually the average number of energy eigenstates within momentum sectors. Finally, we emphasize that $\lambda$ is obtained differently, in the full Hilbert space and in the thermodynamic limit.

\subsection{Set of orthonormal local operators \label{sec:basis}}

 \begin{figure}[!t]
     \centering
     \includegraphics[width=1\linewidth]{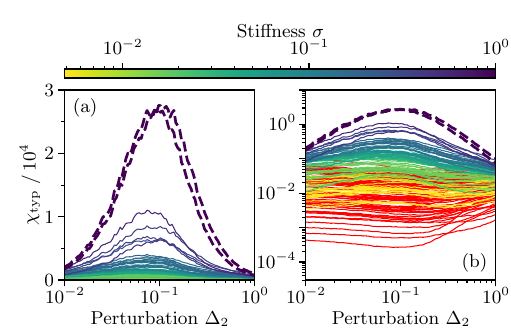}
     \caption{Typical fidelity susceptibility, $\chi_{\rm typ}$, as a function of integrability-breaking strength, $\Delta_2$, for the operators $O^{\alpha}$ characterized by different values of stiffness $\sigma$. Different colors correspond to different values of $\sigma$, as indicated by the colorbar. We consider the system size $L=18$ and the maximal support $M=6$. LIOMs with $\sigma=1$ are marked with a dashed line. Operators with $\sigma\simeq 0$ (within numerical precision) are marked in red. In panel (a), the vertical axis is linear, while in panel (b), it is logarithmic.}
     \label{fig:fs_vs_d2}
 \end{figure}

\begin{figure}[!t]
    \centering
    \includegraphics[width=1\linewidth]{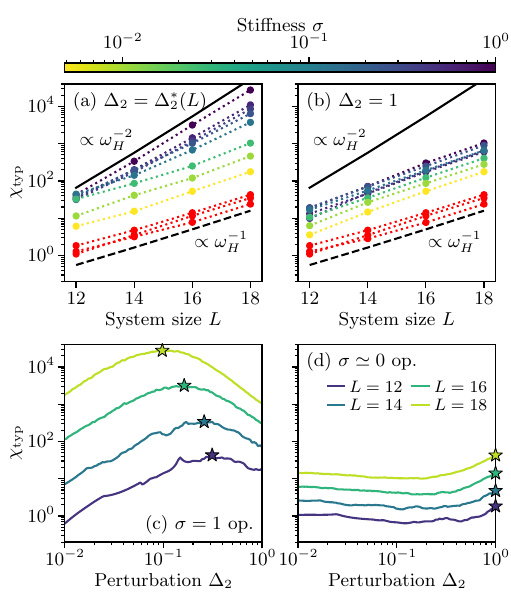}
    \caption{The finite-size scaling of $\chi_\text{typ}$ (a)~at its peak ($\Delta_2=\Delta_2^*$) and (b)~in the ergodic regime ($\Delta_2=1$). Different colors correspond to operators with different $\sigma$. The latter were calculated at $\Delta_2=0$ for $L=18$, as indicated by the colorbar. The solid and dashed lines mark $\chi_\text{typ}\sim \omega_H^{-2}$ and $\chi_\text{typ}\sim \omega_H^{-1}$, respectively. (c) $\chi_\text{typ}$ versus $\Delta_2$ for a LIOM, i.e., for $\sigma=1$, and (d) $\chi_\text{typ}$ versus $\Delta_2$ for an operator with $\sigma\simeq0$ (within numerical precision). We consider $L=12,...,18$. Stars mark the maximal values in the interval.}
    \label{fig:scaling}
\end{figure}

\begin{figure*}[!t]
    \centering
    \includegraphics[width=1\linewidth]{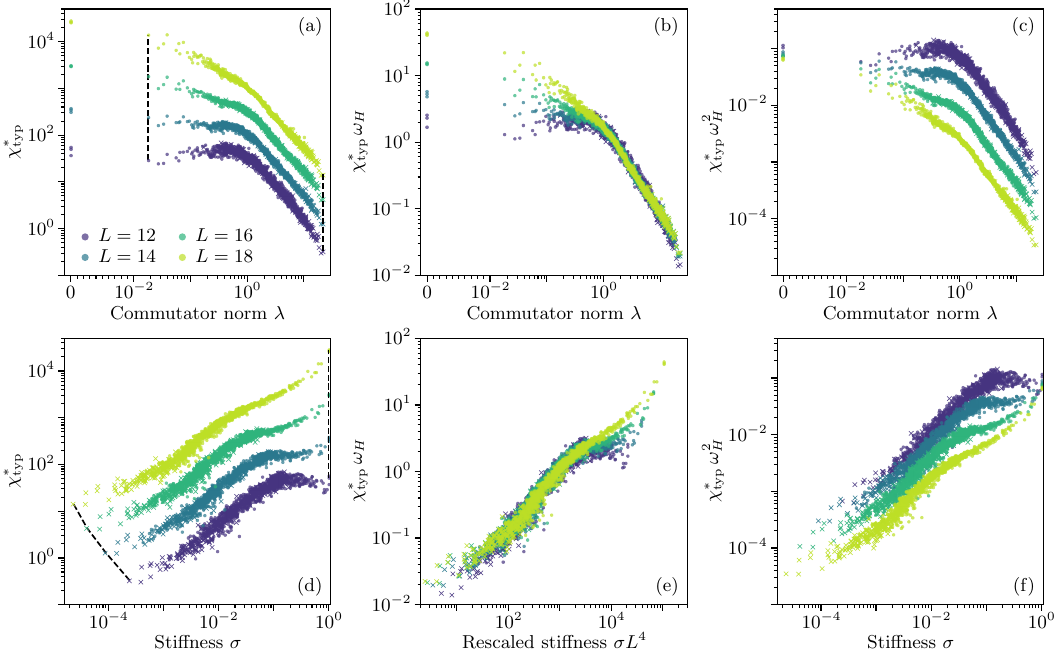}
    \caption{The peak height, $\chi_\text{typ}^*$, plotted versus (a--c)~the commutator norm, $\lambda$, and (d--f)~the stiffness, $\sigma$. The peak height is rescaled by $\omega_H$ in panels (b,e) and by $\omega_H^2$ in panels (c,f), while the stiffness is rescaled by $L^4$ in panel (e). We consider (even) system sizes $L=12,\ldots,18$ and the maximal support $M=8$. In all cases, we set $\omega_H=\sqrt{L}/\mathcal{D}$ for simplicity. Circles (crosses) denote operators that develop a peak (do not develop a peak). Each dashed line in (a) and (d) connects the data points corresponding to the same operator.}
    \label{fig:fid_peak}
\end{figure*}

We plot the typical fidelity susceptibility, $\chi_\text{typ}$, against the integrability-breaking strength, $\Delta_2$, in Fig.~\ref{fig:fs_vs_d2}. We consider a single system size $L=18$ and the set of local operators with the maximal support $M=6$. Different colors correspond to different values of stiffness, $\sigma$, calculated at $\Delta_2=0$. LIOMs, which have $\sigma=1$, are marked with a dashed line. Operators with $\sigma\simeq0$ (within numerical precision) are marked in red. Their negligible stiffness is a consequence of the symmetries they exhibit, as discussed in detail in Sec.~\ref{sec:zeros}. In panel (a), the vertical axis is linear, while in panel (b), it is logarithmic.

It is already clear from Fig.~\ref{fig:fs_vs_d2} that the peak height of the typical fidelity susceptibility, which we will denote as $\chi_\text{typ}^*$ from now on, is related to the stiffness, as the two increase together. We further confirm this observation in Fig.~\ref{fig:scaling}(a), where we plot the finite-size scaling of $\chi_\text{typ}^*$ for a few selected operators. As before, different colors correspond to different $\sigma$, which were calculated at $\Delta_2=0$ for $L=18$. Let us recall that for local operators from the considered set, $\sigma$ is either equal to one and independent of system size (a LIOM) or decays at least polynomially with system size (a local operator orthogonal to all LIOMs). The solid line indicates the maximal divergence, $\chi_\text{typ}\sim \omega_H^{-2}$, while the dashed line indicates the divergence expected in the ergodic system, $\chi_\text{typ}\sim \omega_H^{-1}$. We not only find operators for which $\chi_\text{typ}^*$ closely follows these limiting behaviors, but also operators for which it exhibits an intermediate one. 

In Fig.~\ref{fig:scaling}(b), we verify that the typical fidelity susceptibility for the same operators as those in Fig.~\ref{fig:scaling}(a) is consistent with the scaling $\chi_\text{typ}\sim\omega_H^{-1}$ deep in the ergodic regime ($\Delta_2=1$). Additionally, for a particular case of a LIOM, we demonstrate how the peak position, which is marked with a star, evolves with the system size in Fig.~\ref{fig:scaling}(c). We consider (even) system sizes $L=12,\ldots,18$. In agreement with expectations, it shifts towards smaller $\Delta_2$ with increasing $L$. For completeness, we present similar numerical results for an operator with $\sigma\simeq 0$ (within numerical precision) in Fig.~\ref{fig:scaling}(d). For this case, the typical fidelity susceptibility does not exhibit a peak. Its maximal value is attained at the edge of the considered interval, i.e., $\Delta_2=1$.

All the observations from Figs.~\ref{fig:fs_vs_d2} and~\ref{fig:scaling} are confirmed by the numerical results presented in Fig.~\ref{fig:fid_peak}, which provides a detailed analysis of the complete set of operators with the maximal support $M=8$. We consider (even) system sizes $L=12,\ldots,18$. In the top row of Fig.~\ref{fig:fid_peak}, we plot the peak height, $\chi_\text{typ}^*$, versus the commutator norm, $\lambda$. Specifically, we show $\chi_\text{typ}^*$ versus $\lambda$ in Fig.~\ref{fig:fid_peak}(a),  $\chi_\text{typ}^* \omega_H$ versus $\lambda$ in Fig.~\ref{fig:fid_peak}(b), and $\chi_\text{typ}^* \omega_H^2$ versus $\lambda$ in Fig.~\ref{fig:fid_peak}(c). We note that $\lambda$ has been established from the thermodynamic limit calculations and, so, it is independent of system size for a given operator. Results in Figs.~\ref{fig:fid_peak}(a)-\ref{fig:fid_peak}(c) allow us to reveal two distinct behaviors. For $\lambda\gtrsim 1$, the peak height seems to follow the scaling expected in the ergodic regime, $\chi_\text{typ}^*\sim \omega_H^{-1}$, see Fig.~\ref{fig:fid_peak}(b). Simultaneously, for $0<\lambda\lesssim 1$, the peak height exhibits intermediate scaling, faster than $\chi_\text{typ}^*\sim \omega_H^{-1}$ and slower than $\chi_\text{typ}^*\sim \omega_H^{-2}$, see Fig.~\ref{fig:fid_peak}(c).
The boundary separating these two behaviors appears to shift towards larger $\lambda$ with increasing $L$. Nevertheless, the latter cannot be unambiguously confirmed from the numerical results.

The bottom row of Fig.~\ref{fig:fid_peak} shows the same numerical results as the top row, but plotted versus the stiffness, $\sigma$. Specifically, we show $\chi_\text{typ}^*$ versus $\sigma$ in Fig.~\ref{fig:fid_peak}(d),  $\chi_\text{typ}^* \omega_H$ versus $\sigma L^4$ in Fig.~\ref{fig:fid_peak}(e), and $\chi_\text{typ}^* \omega_H^2$ versus $\sigma$ in Fig.~\ref{fig:fid_peak}(f). It appears that larger $\sigma$ generally leads to larger $\chi_\text{typ}^*$, consistent with both the relation in Eq.~\eqref{eq:final}, expected to hold in the limit $\sigma\to 1$, and the bound in Eq.~\eqref{eq:bound}, applicable to all $\sigma$. Moreover, for $L\gtrsim 18$, the dependence of $\chi_\text{typ}^*$ on $\sigma$ appears to become linear. In Sec.~\ref{sec:generic}, we demonstrate that it is well approximated by Eq.~\eqref{eq:final}. Such a linear dependence does not contradict the different possible finite-size scalings of the peak, since $\sigma$ itself depends on $L$, except for LIOMs, for which $\sigma=1$. Therefore, as the system size increases, the data point corresponding to a given operator with $\sigma<1$ shifts toward smaller values of $\sigma$ along the horizontal axis.

Finally, in Fig.~\ref{fig:fid_peak}(e), we observe a scaling collapse of $\chi_\text{typ}^*\omega_H$ when plotted versus $\sigma L^4$, consistent with a negligible peak or its complete absence.
The prefactor $L^4$ was found empirically to produce the best data collapse. Note that, for the available system sizes $L\le 18$, a polynomial decay $\sigma \propto 1/L^4$ or faster is indistinguishable from an exponential decay.

\subsection{Generic local operators \label{sec:generic}}

To strengthen our observations, we also consider linear combinations of local operators from the orthonormal set studied in the previous sections:
\begin{equation} \label{def:Otilde}
\tilde{O}= a_\alpha O^\alpha  +a_\beta O^\beta,
\end{equation}
where $a_\alpha$ and $a_\beta$ are real numbers satisfying the condition $a_\alpha^2+a_\beta^2=1$.
We parametrize them as $a_{\alpha} = \cos(\varphi), \, a_{\beta} = \sin(\varphi)$, with $\varphi \in [ 0, \pi/2 ]$.
Note that this allows us to generate, among others, local operators that are not LIOMs but have a finite $\sigma<1$ in the thermodynamic limit. Generic local operators studied in the context of nonequilibrium quantum dynamics, which are not intentionally constructed to be orthogonal to all LIOMs, have this property. According to our derivation, they are expected to exhibit maximal chaos, see the discussion in Sec.~\ref{sec:fidelity}.

\begin{figure}[!t]
    \centering
    \includegraphics[width=1.0\linewidth]{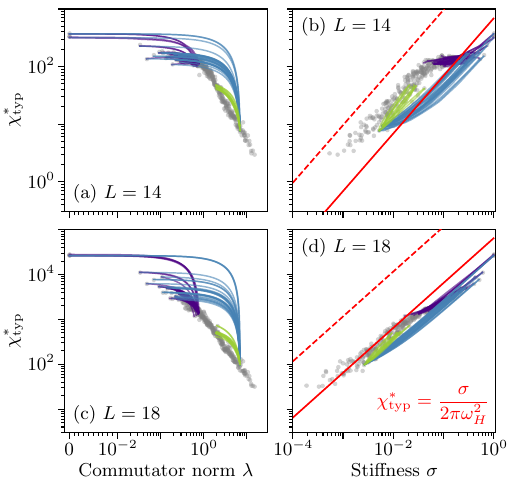}
    \caption{The peak height, $\chi_\text{typ}^*$, versus (a,c)~the commutator norm, $\lambda$, and (b,d)~stiffness, $\sigma$. We consider (a,b)~$L=14$ and (c,d)~$L=18$. We focus on $M=7$. Semitransparent (grey) symbols mark the results for the operators $O$ from Eq.~\eqref{def:Oalpha}, i.e., for the orthonormal set of operators. They are identical to the results in Fig.~\ref{fig:fid_peak}. Lines mark the results for the operators $\tilde{O}$ from Eq.~\eqref{def:Otilde}. Green (purple) lines correspond to the cases of $\tilde{O}$ when both $O_\alpha$ and $O_\beta$ have large (small) $\lambda$. Simultaneously, blue lines correspond to the cases when $O_\alpha$ has a large $\lambda$, while $O_\beta$ has a small $\lambda$. Solid red lines in (b,d) denote the predicted behavior from Eq.~\eqref{eq:final} for a fixed value of a coefficient $a=1$,
    while dashed lines in (b,d) correspond to the bound from Eq.~\eqref{eq:bound} with $a=1$ and $b=1$.
    }
    \label{fig:placeholder}
\end{figure}

For these linear combinations, $\tilde{O}$ from Eq.~\eqref{def:Otilde}, we plot the peak height, $\chi_\text{typ}^*$, versus the commutator norm, $\lambda$, and stiffness, $\sigma$, in the left and right columns of Fig.~\ref{fig:placeholder}, respectively. We consider the system size $L=14$ ($L=18$) in the top (bottom) row of Fig.~\ref{fig:placeholder}. We focus on support $M=7$. Semitransparent grey symbols mark the numerical results for the operators $O$ from Eq.~\eqref{def:Oalpha}, while lines mark the results for the operators $\tilde{O}$ from Eq.~\eqref{def:Otilde}.

There are three types of $\tilde{O}$ in Fig.~\ref{fig:placeholder}, distinguished by the target $\lambda$ ranges of $(O^{\alpha}, O^{\beta})$ and marked with different colors. For each type, we randomly select $50$ pairs $(O^{\alpha}, O^{\beta})$, and for each pair, a single line in Fig.~\ref{fig:placeholder} is obtained by taking $50$ linearly spaced values of $\varphi \in [0, \pi/2]$. Green (purple) lines correspond to the cases when both $O_\alpha$ and $O_\beta$ have large $\lambda \gg 1$ (small $\lambda \ll 1$). These lines lie close to the semitransparent grey symbols, indicating that the resulting local operators $\tilde{O}$ retain either the scaling of the peak expected in the ergodic regime, $\chi_\text{typ}^*\sim\omega_H^{-1}$, or a faster scaling, $\chi_\text{typ}^*\sim\omega_H^{-\kappa}$ with $\kappa\in(1,2]$.

We now move our attention to blue lines, which correspond to the cases when $O_\alpha$ has a large $\lambda\gg 1$, while $O_\beta$ has a small $\lambda\ll 1$. It is apparent from Figs.~\ref{fig:placeholder}(a) and~\ref{fig:placeholder}(c) that the peak height is no longer determined by the commutator norm. Specifically, even a small admixture of a local operator $O_\beta$ with $\lambda\ll 1$ induces a faster scaling $\chi_\text{typ}^*\sim\omega_H^{-\kappa}$ with $\kappa\in(1,2]$.
Simultaneously, as observed from Figs.~\ref{fig:placeholder}(b) and~\ref{fig:placeholder}(d), these blue lines closely follow the relation from Eq.~\eqref{eq:final}, which is marked by a red line for $a=1$. The agreement is observed for $\sigma\ll 1$ and all considered system sizes, even before the peak hight, $\chi_\text{typ}^*$, becomes a linear function of stiffness, $\sigma$. It can be understood from the fact that $O_\beta$ is expected to have a dominant projection onto a single conserved quantity (a LIOM or a product of LIOMs), and the same holds for the linear combination $\tilde{O}$.

\subsection{Local operators with zero stiffness}
\label{sec:zeros}

When investigating the orthonormal set from Eq.~\eqref{def:Oalpha}, we encountered many local operators whose stiffness was zero within numerical precision. In this section, we provide an analytical argument showing that they arise from symmetry constraints, and then analyze their properties.

We first note that the investigated Hamiltonian, $H$, can be represented by a real matrix in the real-space basis, $\{\ket{f}\}$, where each state corresponds to a configuration of spins. This follows from its time-reversal symmetry. Consequently, the energy eigenstates, $\{\ket{n}\}$, can be obtained via an orthogonal transformation and have real components, i.e., $\langle f|n\rangle \in \mathbb{R}$. Let us now consider operators, $O$, that are represented by purely imaginary matrices in the real-space basis. Such matrices remain purely imaginary when transformed to the energy eigenbasis, which implies the vanishing of their diagonal matrix elements, $\langle n|{O}|n\rangle = -\langle n|{O}|n\rangle^*=0$.

Simultaneously, the investigated Hamiltonian is translationally invariant, so its eigenstates can be chosen to have well-defined momentum, $\{\ket{n,k}\}$. At the same time, it is even under the mirror symmetry, $\mathcal{M}$. Momentum is odd under this symmetry, so that $\mathcal{M}\ket{n,k} \sim \ket{n,-k}$ for all $k \neq 0, \pi$. The latter are excluded from our numerical simulations. Consequently, the energy spectrum is doubly degenerate:
\begin{equation}
    H{\cal M} \ket{n,k}={\cal M}H \ket{n,k}=E_{n,k}{\cal M} \ket{n,k}.
\end{equation}
It is straightforward to demonstrate that these degenerate pairs correspond to Kramers pairs.

In each degenerate subspace, the considered operators, ${O}$, can be written as the matrices $\mathcal{{O}}=A\sigma^y$, where $A \in \mathbb{R}$ and $\sigma^y$ denotes the Pauli matrix. Additionally, if ${O}$ is translationally invariant, it has zero matrix elements between states with different momenta. This implies that the matrix $\mathcal{{O}}=A\sigma^y$ becomes diagonal in the basis $\{\ket{n,k}, \ket{n,-k}\}$ with the diagonal elements $A$ and $-A$. Finally, if the considered operator, ${O}$, is even under the mirror symmetry, we can write for all $k \neq 0, \pi$ that
\begin{equation}
\begin{split}
    A & =\langle n,k|{O}|n,k\rangle=\langle n,k|\mathcal{M}^{-1}(\mathcal{M}{O}\mathcal{M}^{-1})\mathcal{M}|n,k\rangle\\
    & = \langle n,-k|{O}|n,-k\rangle=-A,
    \end{split}
\end{equation}
which in turn enforces $\langle n,k|{O}|n,k\rangle=0$. In the above equation, we have used that the mirror symmetry is unitary, i.e., $\mathcal{M}^\dagger = \mathcal{M}^{-1}$. As a result, the stiffness is exactly zero for all ${O}$ that are purely imaginary in the real-space basis, translationally invariant, and even under mirror symmetry.

\begin{figure}
    \centering
    \includegraphics[width=1.0\linewidth]{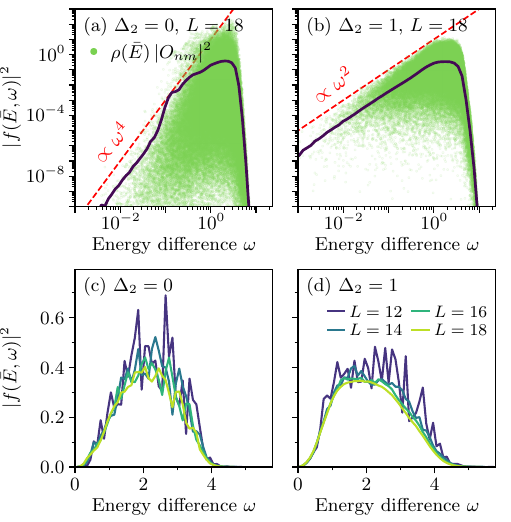}
    \caption{The envelope function, $|f(\bar{E},\omega)|^2$, of the operator with zero stiffness from Eq.~\eqref{eq:zero_mode}. We plot it at (a,c) the integrable point ($\Delta_2=0$) and (b,d) in the ergodic regime ($\Delta_2=1$). We use logarithmic scales in (a,b), and linear scales in (c,d). Green points in (a,b) mark the rescaled offdiagonal matrix elements, $\rho(\bar{E})|O_{nm}|^2$, for $L=18$. Different colors in (c,d) correspond to different system sizes, $L=12,\ldots,18$.
    The dashed lines in (a) and (b) serve as guides for the eye.}
    \label{fig:zero_mode}
\end{figure}

The simplest operator satisfying the conditions described above is 
\begin{equation}
\begin{split}
   {O} &= i\sum_l(S_l^z S_{l+1}^+ S_{l+2}^- + S_l^- S_{l+1}^+ S_{l+2}^z)\\
    &- i\sum_l(S_l^z S_{l+1}^- S_{l+2}^+ + S_l^+ S_{l+1}^- S_{l+2}^z),
    \label{eq:zero_mode}
\end{split}
\end{equation}
which is similar to the operators studied in the context of spin chirality~\cite{Frahm_1997,Wietek_2017,Wei_2024}.
Although its typical fidelity susceptibility, $\chi_\text{typ}$, does not develop a peak, it is still useful to describe the behavior of its envelope function, $|f(\bar{E},\omega)|^2$.
In the left and right columns of Fig.~\ref{fig:zero_mode}, we plot $|f(\bar{E},\omega)|^2$ at the integrable point ($\Delta_2=0$) and ergodic regime ($\Delta_2=1$), respectively. In both cases, $|f(\bar{E},\omega)|^2$ vanishes as $\omega\rightarrow 0$. This has two consequences. 

First, the envelope function is continuous at $\omega=0$, allowing the matrix elements of $O$ to satisfy the ETH ansatz. A comment is in order here. The ETH ansatz is commonly understood to predict that fluctuations of both diagonal and off-diagonal matrix elements decay exponentially with system size, with the former measured relative to the so-called structure function. More precisely, however, these fluctuations are described by Eq.~\eqref{ETH} extended to the $n=m$ case, with the exponential suppression arising from the inverse density of states, i.e., $\rho(E_{nm})^{-1}$. Importantly, $f(E_{nm},0)=0$, although corresponding to the absence of fluctuations of diagonal matrix elements, is still consistent with Eq.~\eqref{ETH}. We therefore regard such behavior as satisfying the ETH ansatz. It is also worth emphasizing that $|f(\bar{E},\omega)|^2$ appears to decay polynomially with decreasing $\omega$ in the ergodic regime [Fig.~\ref{fig:zero_mode}(b)], whereas it decays visibly faster and may even exhibit a gap at the integrable point [Fig.~\ref{fig:zero_mode}(a)]. Additionally, the fluctuations of off-diagonal matrix elements of $O$ are larger at the integrable point, but decay exponentially with system size in both cases.

Second, $|f(\bar{E},\omega)|^2$ develops a maximum at $\omega\neq 0$. As shown in the bottom row of Fig.~\ref{fig:zero_mode}, the position of this maximum does not appear to shift towards smaller $\omega$ as the system size increases, at least for considered $L=12, ..., 18$. Since the (connected) autocorrelation function is the Fourier transform of $|f(\bar{E},\omega)|^2$, it exhibits (approximately) periodic oscillations. Although the amplitude of these oscillations decreases with time, it does not appear to decrease with system size.

\section{Conclusions}
\label{sec:conclusions}

In this manuscript, we studied the behavior of the typical fidelity susceptibility, $\chi_\text{typ}$, across the integrability-breaking transition. This quantity measures the sensitivity of energy eigenstates to perturbations. Previous studies have shown that it can develop a peak signaling the breaking of integrability and the onset of ergodic behavior~\cite{PhysRevB.109.195162,Abdelshafy_2025,Kim_2026}. It has also been argued that this peak exhibits maximal divergence for integrability-preserving perturbations, whereas it is absent for integrability-breaking ones~\cite{PhysRevX.10.041017}. However, these two classes of perturbations have never been sharply defined, and examples that challenge this classification have also been reported in the literature~\cite{LeBlond_2021,Lisiecki_2025}.

We provided a simple argument that the peak height, $\chi_\text{typ}^*$, is determined by the stiffness, $\sigma$, in the integrable limit. For large systems, $\chi_\text{typ}^*$ is given by the expression from Eq.~\eqref{eq:final}. This argument assumes that breaking integrability leads to the broadening of the Drude-like peak and that the peak in the fidelity susceptibility emerges when this broadening becomes comparable to the typical level spacing. The stiffness, defined as the mean of squared diagonal matrix elements, see Eq.~\eqref{eq:stiff}, is determined by the projection of the perturbation onto local integrals of motion (LIOMs), as demonstrated in Eq.~\eqref{eq:stiff2}. As a result, local perturbations with a non-vanishing projection onto LIOMs exhibit maximal chaos, while those orthogonal to all LIOMs can exhibit an arbitrary intermediate scaling of $\chi_\text{typ}^*$, ranging from the maximal divergence to the absence of a peak.

We supported our argument with numerical simulations of the XXZ spin chain with next-nearest-neighbor interactions. Rather than studying randomly selected local perturbations, we generated an orthonormal set of operators using the method described in Ref.~\cite{Paw_owski_2025}. This set included all LIOMs and other local perturbations up to a given support, and we also analyzed various linear combinations of its elements. The latter include generic local operators studied in the context of nonequilibrium quantum dynamics, which are not LIOMs but retain a finite projection onto them in the thermodynamic limit.

Finally, we investigated local perturbations whose stiffness is
zero within numerical precision. We demonstrated that $\sigma=0$ is enforced by the symmetry constraints. We showed that the matrix elements of such local perturbations satisfy the ETH ansatz both at the integrable point and in the ergodic regime, see the discussion in Sec.~\ref{sec:zeros}. Additionally, their autocorrelation functions exhibit (approximately) periodic oscillations with amplitude decreasing with time but not system size.

\begin{acknowledgments}

We acknowledge discussions with A.~Polkovnikov. 
J.~P. acknowledges support from the National Science Centre (NCN), Poland via project 2023/49/N/ST3/01033.
P.~Ł. acknowledges support from the National Science Centre (NCN), Poland via project 2025/59/D/ST3/00229.
L.~V. acknowledges support from the Slovenian Research and Innovation Agency (ARIS), Research core funding Grants No.~P1-0044, N1-0273 and J1-50005.

\section*{Data Availability}

The data that support the findings of this article are openly available~\cite{data_repository}.

\end{acknowledgments}

\bibliographystyle{biblev1}
\bibliography{manupert}

\end{document}